\documentclass[%
 reprint,
 amsmath,amssymb,
 aps,
]{revtex4-1}

\usepackage{graphicx}
\usepackage{dcolumn}
\usepackage{bm}

\begin{document}

\preprint{APS/123-QED}

\title{Inverse Edelstein effect induced by magnon-phonon coupling}

\author{Mingran Xu}
\affiliation{Institute for Solid State Physics, University of Tokyo, 5-1-5 Kashiwanoha, Kashiwa, Chiba, 277-8581, Japan}
\author{Jorge Puebla}
\email{jorgeluis.pueblanunez@riken.jp}
\affiliation{CEMS, RIKEN, Saitama, 351-0198, Japan}
\author{Florent Auvray}
\affiliation{Institute for Solid State Physics, University of Tokyo, 5-1-5 Kashiwanoha, Kashiwa, Chiba, 277-8581, Japan}
\author{Bivas Rana}
\affiliation{CEMS, RIKEN, Saitama, 351-0198, Japan}
\author{Kouta Kondou}
\affiliation{CEMS, RIKEN, Saitama, 351-0198, Japan}
\author{Yoshichika Otani}
\email{yotani@issp.u-tokyo.ac.jp}
\affiliation{Institute for Solid State Physics, University of Tokyo, 5-1-5 Kashiwanoha, Kashiwa, Chiba, 277-8581, Japan}
\affiliation{CEMS, RIKEN, Saitama, 351-0198, Japan}

\date{\today}

\begin{abstract}
We demonstrate a spin to charge current conversion via magnon-phonon coupling and inverse Edelstein effect on the hybrid device Ni/Cu(Ag)/Bi$_{2}$O$_{3}$. The generation of spin current ($J_{s}\approx 10^{8}A/m^{2}$) due to magnon - phonon coupling reveals the viability of acoustic spin pumping as mechanism for the development of spintronic devices. A full in-plane magnetic field angle dependence of the power absorption and a combination of longitudinal and transverse voltage detection reveals the symmetric and asymmetric components of the inverse Edelstein effect voltage induced by Rayleigh type surface acoustic waves. While the symmetric components are well studied, asymmetric components are widely unexplored. We assign the asymmetric contributions to the interference between longitudinal and shear waves and an anisotropic charge distribution in our hybrid device.
\end{abstract}

\pacs{Valid PACS appear here}
\maketitle


Methods for generation of spin current and its conversion to electrical charge current has been vigorously studied in recent years, particularly based on physical phenomena at the nanoscale \cite{Otani}. In the presence of magnetic materials, spin currents can be generated by magnon-photon or magnon-phonon coupling. Magnon-phonon coupling can be achieved by passing surface acoustic waves (SAW) across ferromagnetic layers due to magnetoelastic effect \cite{Weiler1}. Periodic elastic deformation of ferromagnetic film by SAW drives precessional magnetization dynamics such as ferromagnetic resonance (FMR). This process is known as acoustic-FMR (A-FMR), and is analogous to the most common FMR which is driven by electromagnetic waves (photons). Both, FMR and A-FMR are used as mechanism for generation of spin current which can be injected into an adjacent nonmagnetic layers, a process better known as spin pumping \cite{Yaro, Weiler2}. The generated spin current is usually converted to electrical charge current by inverse spin Hall effect (ISHE) \cite{Mosendz}. Alternatively, recent reports showed efficient spin to charge current conversion at interfaces with spatial inversion asymmetry between two nonmagnetic materials \cite{Karube, Rojas}. Spatial inversion asymmetry induces a built-in electric potential and spin orbit coupling at surfaces and interfaces, the so-called Rashba spin orbit coupling \cite{Ando}. Here, the spin to charge conversion mechanism is known as inverse Edelstein effect (IEE). In this letter, we demonstrate spin to charge conversion in a hybrid device which combines magnon-phonon coupling via SAW and inverse Edelstein effect (IEE). 

Our hybrid devices consist of  Ni(10nm)/Cu(20nm)/Bi$_{2}$\\O$_{3}$(20nm) and Ni(10nm)/Ag(20nm)/Bi$_{2}$O$_{3}$(20nm) trilayer structure deposited on LiNbO$_{3}$ substrates at the center of a pair of Ti(5nm)/Au(20nm) interdigital transducer (IDT). By applying RF voltage at the input IDT, Rayleigh type SAWs are launched along x-axis (see Fig. 1(a)), driving a time dependent strain field $\varepsilon(t)$ in the lattice, inducing a time varying contribution to the magnetic anisotropy in the ferromagnetic layer (Ni) via magnetoelastic effect \cite{Weiler1}. As schematically demonstrate in Figure 1 (b), in the presence of a static magnetic field $\textbf{H}$, the SAW induced anisotropy field excites a processional motion of magnetization $\textbf{M}$ around $\textbf{H}$ direction, enabling ferromagnetic resonance in Ni and injecting spin current into the adjacent nonmagnetic layer, Cu (Ag). Then, due to the inversion spatial symmetry breaking at the interface between Cu (Ag) and Bi$_{2}$O$_{3}$ \cite{Karube, Puebla, Kim}, the system converts spin current into charge current via IEE, inducing a magnetic field dependent voltage $V$, detected in longitudinal and transverse geometries.

The spin current generated by SAW via magnon-phonon coupling strongly depends in the absorption of acoustic waves in the ferromagnetic layer. We first characterize the absorption of SAW by using a vector network analyzer (VNA), while varying the in-plane external magnetic field angle. Absorption of SAW is characterized by systematically measuring the transmission coefficient $\overline{\lvert S_{21} \rvert}$ as function of in-plane static magnetic field magnitude and angle $\theta$, as illustrated in Figure 1(a). Figure 2(a) and 2(b) show the magnetic field angle dependence ($\theta$) at the resonance peak  of the absorption of SAW into FMR excitation in Ni. We observe a four fold butterfly shape signal with first maximum located at $\theta=45^{\circ}$. Previous reports showed a four fold symmetric dependence of SAW absorption with the angle formed between SAW $\textbf{k}$-wavevector and in-plane magnetic field, having maximum absorption at $45^{\circ}$, and it is considered as the fingerprint of acoustic-FMR \cite{Weiler1, Dreher}. Asymmetry between top and bottom side of the four fold symmetry indicates the non-negligible interference of magnetoelastic couplings of shear waves and longitudinal waves (Fig. 2(b)). Magnetization dynamics induced by Rayleigh SAW are composed by longitudinal wave and shear wave motion components perpendicular to each other ($\pi /2$ dephasing), forming ellipsoidal oscillations. When the SAW direction vector \textbf{k} is reversed, the ellipsoidal motion is reversed, which is accompanied by a sign change of the shear strain component $\varepsilon_{zx0}$. Magnetoelastic coupling by interference of longitudinal ($\varepsilon_{xx0}$) and shear wave ($\varepsilon_{zx0}$) can be described by Eq. (1), which directly elucidates the excitation efficiency of FMR, which is proportional to the SAW absorption $\Delta P(\theta)\propto[\mu_{0}h_{rf}(\theta)]^{2}$. 

\begin{figure}[t!]
\begin{center}
\includegraphics[width=1.0\columnwidth, height=10.0cm, keepaspectratio]{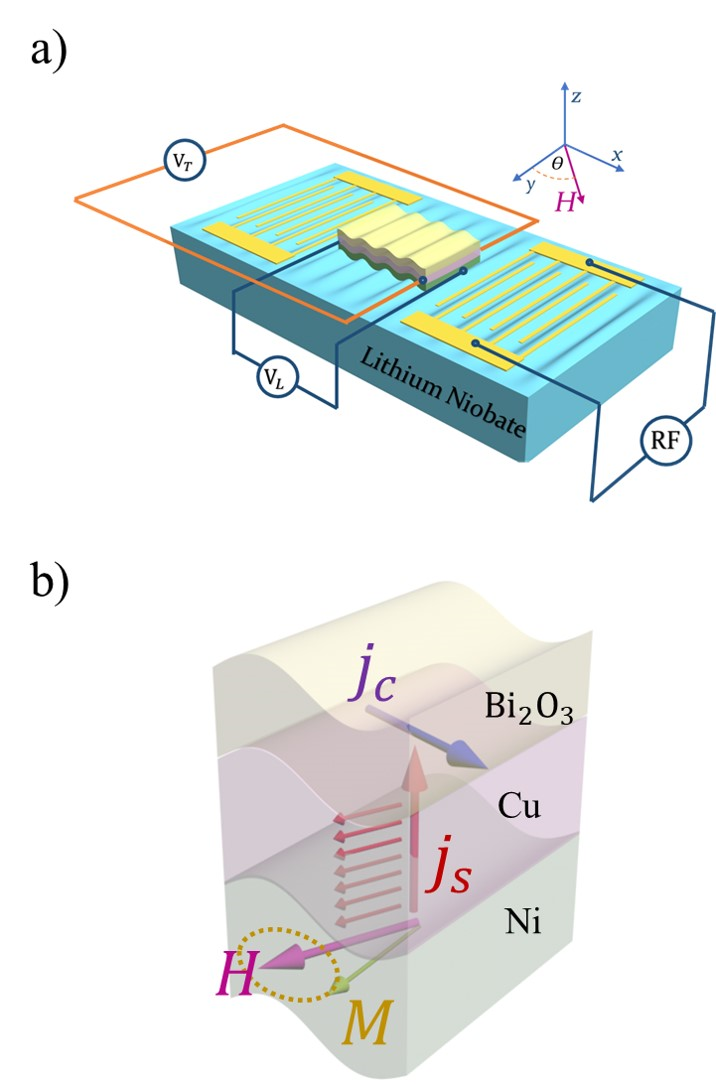}
\caption{(a): Illustration of experiment setup. The surface acoustic waves are generated by applying RF voltage on interdigital transducers and the longitudinal voltage and transverse voltage are measured while applying an external magnetic field \textbf{H} at an angle $\theta$. (b): Schematics of acoustic spin pumping mechanism and spin to charge conversion via IEE.}
\label{fig:1}
\end{center}
\end{figure}

\begin{figure}[t!]
\begin{center}
\includegraphics[width=1.0\columnwidth, height=12.0cm, keepaspectratio]{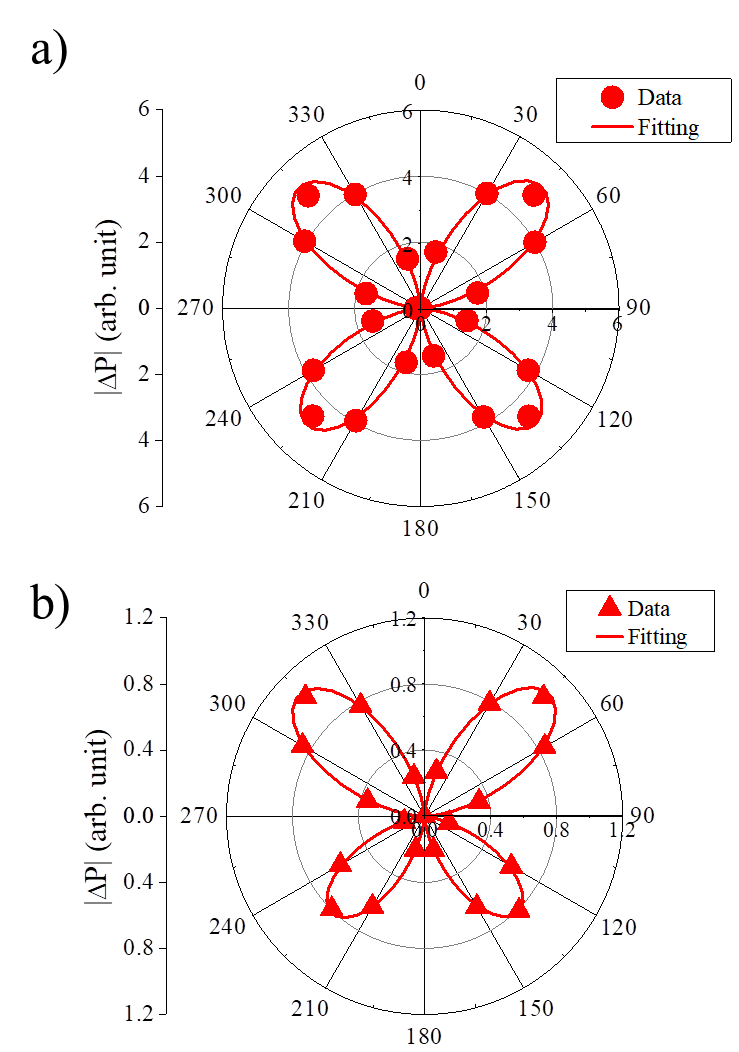}
\end{center}
\caption{Polar graph of damping of SAW due to acoustically excited FMR measured on (a): Ni/Cu/Bi$_{2}$O$_{3}$ and (b): Ni/Ag/Bi$_{2}$O$_{3}$, varying applied magnetic field angle $\theta$.}
\label{fig:main}
\end{figure}

\begin{figure}[b!]
\begin{center}
\includegraphics[width=0.8\columnwidth]{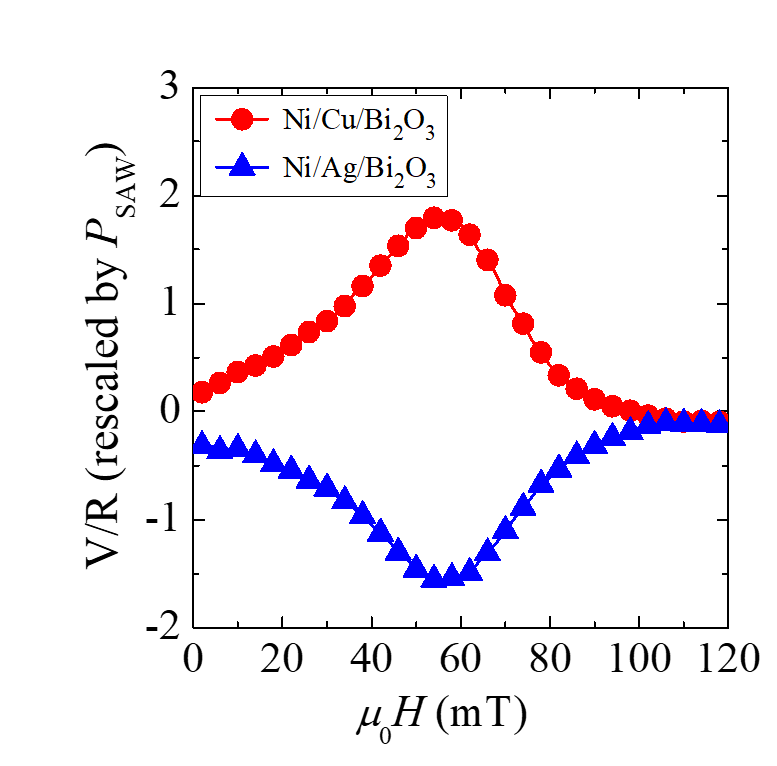}
\end{center}
\caption{Longitudinal signal measured on Ni/Cu/Bi$_{2}$O$_{3}$ (red circles) and Ni/Ag/Bi$_{2}$O$_{3}$ (blue triangles) devices depending on the magnetic field amplitude at $\theta=240^{\circ}$.}
\label{fig:2}
\end{figure}

\begin{figure*}[t!]
\includegraphics[width=2.0\columnwidth]{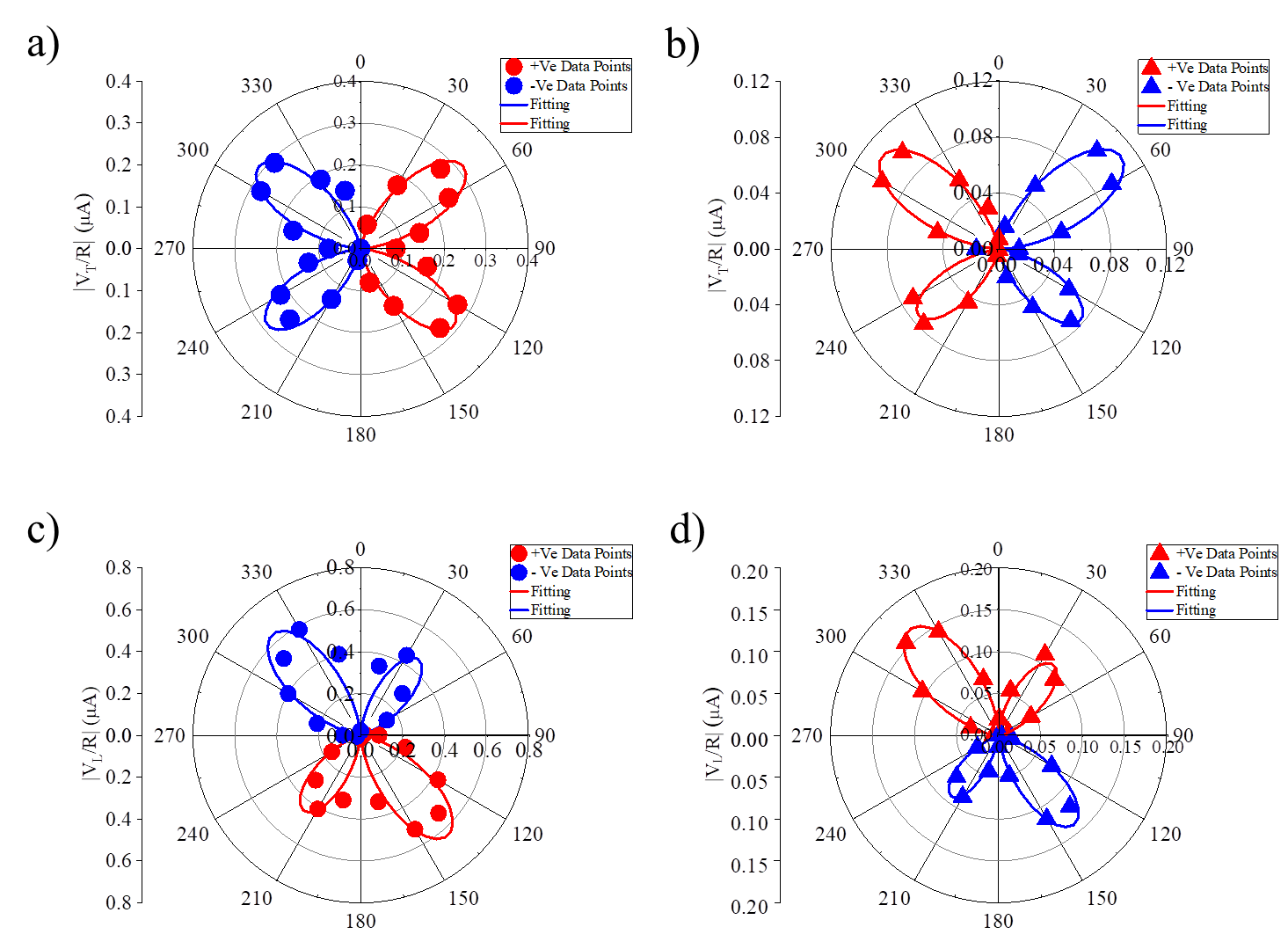}
\caption{Polar graph of transverse signal measured on (a): Ni/Cu/Bi$_{2}$O$_{3}$ and (b): Ni/Ag/Bi$_{2}$O$_{3}$, varying applied external magnetic field angle $\theta$. Polar graph of longitudinal signal measured on (c): Ni/Cu/Bi$_{2}$O$_{3}$ and (d): Ni/Ag/Bi$_{2}$O$_{3}$, varying applied external magnetic field angle $\theta$.}
\label{fig:main}
\end{figure*}

\begin{equation}
\begin{split}
[\mu_{0}h_{rf}(\theta)]^{2}=[b_{1}\varepsilon_{xx0}sin\theta cos\theta+2b_{2}\varepsilon_{zx0}sin\theta]^{2}
\label{eq:H-NV}
\end{split}
\end{equation}

where, $b_{1(2)}$ is the magneto-elastic coupling constant, $\varepsilon_{xx0(zx0)}$ represents the longitudinal strain (shear strain) induced by SAW, and $\theta$ refers to in-plane static magnetic field orientation as depicted in Figure 1(a). Recently, the microscopic origin of this asymmetric mechanism has been reported and it has been ascribed to simultaneous breaking of time reversal and spatial inversion symmetries \cite{Onose}. Both, time reversal and spatial inversion symmetry are broken at our hybrid device. The resultant asymmetry is particularly evident for Ni/Ag/Bi$_{2}$O$_{3}$ in Fig. 2(b). Ratio of longitudinal and transverse magnetoelastic coupling in Ag is larger than Cu, inducing larger anisotropies \cite{Nazar}. 

Driving of acoustic-FMR produces spin current which can be injected to adjacent nonmagnetic layers, and converted to charge current when spin orbit interaction exists. Figure 3 shows the magnetic field dependent  IEE charge current detected in Ni/Cu/Bi$_{2}$O$_{3}$ (red circles) and Ni/Ag/Bi$_{2}$O$_{3}$ (blue triangles) devices, rescaled by their corresponding SAW power absorption, P$_{SAW}$. The spin to charge conversion at nonmagnetic metal/Bi$_{2}$O$_{3}$ interfaces was first reported by Karube et. al. \cite{Karube}, describing the non-equilibrium spin accumulation induced by spin current injection at this interface, resulting in a shift of Fermi contours in momentum space that generates a flow of charge current in the device. Considering the fact that the ISHE is negligible in our device, the spin pumping signal is mainly consequence of the IEE. Therefore, the opposite sign of the spin pumping signal shown in Fig. 3 reflects the opposite spin momentum locking configuration at Cu/Bi$_{2}$O$_{3}$ and Ag/Bi$_{2}$O$_{3}$ interfaces, corroborating well the observation from electromagnetic wave induced spin pumping and MOKE \cite{Puebla, Tsai}. Spin pumping measurements driven by electromagnetic wave FMR contain an asymmetric component due to classical induction effect, the anisotropic magnetoresistance voltage \cite{Azevedo}. The induction effect is strongly suppressed in acoustic spin pumping measurements. 

Feasibility of acoustic spin pumping was previously reported in a limited range of in-plane magnetic field angles, measured in transverse geometry \cite{Weiler2}. We extend our study to the full in-plane magnetic field angle dependence with voltage detection in both, transverse and longitudinal geometries. For transverse geometry we detect the voltage perpendicular to the SAW wavevector \textbf{k} (V$_{T}$ in Fig. 1(a)), while for the longitudinal geometry we detect the voltage parallel to the SAW wavevector \textbf{k} (V$_{L}$ in Fig. 1(a)). Figure 4(a) and 4(b) show the magnetic field angle dependence ($\theta$) of the transverse IEE voltage V$_{T}$ for Ni/Cu/Bi$_{2}$O$_{3}$ (a) and Ni/Ag/Bi$_{2}$O$_{3}$ (b). Since the V$_{T}$ mainly depends on the excitation efficiency of FMR and the projection angle of the spin current onto the Rashba interface, the top and bottom side asymmetry presented in SAW absorption is also reflected in the transverse voltage. For fitting our data, as the SAW damping in longitudinal direction does not affect the transverse potential in our device, the longitudinal component of $\mu_{0}h_{rf}(\theta)$ has negligible influence in our spin pumping signal, and we can only consider the transverse component of excitation $\lvert \mu_{0}h_{rf} \rvert sin(\theta)$ in (2)
\begin{equation}
V_{T}(\theta)/R=C_{1}[\mu_{0}h_{rf}]^{2}sin(\theta)
\label{eq:TVolt}
\end{equation}

We repeat our acoustic spin pumping measurements now with longitudinal voltage detection. Figures 4(c) and 4(d) show the magnetic field angle dependence ($\theta$) of the longitudinal IEE voltage V$_{L}$ for Ni/Cu/Bi$_{2}$O$_{3}$ (c) and Ni/Ag/Bi$_{2}$O$_{3}$ (d). Due to the damping of SAW in the propagation direction, it is possible to observe influence of the transverse excitation component in the V$_{L}$, adding an amplitude asymmetry in the four fold symmetry. We fit our data by (3)

\begin{equation}
\begin{split}\label{eq:LVolt}
V_{L}(\theta)/R= [\mu_{0}h_{rf}]^{2}[C_{T}sin(\theta)+C_{L}cos(\theta)]
\end{split}
\end{equation}

We also observe an additional asymmetric component in Figs. 4(c) and 4(d). This asymmetric component is described by the non-symmetric charges distribution  at the position of our electrodes, when varying the magnetic field angle. The charge distribution is dictated by the electric field \textbf{E} induced by spin orbit interaction, which is proportional to the flow direction of spin current J$_{s}$ and its spin polarization $\sigma_{s}$, such that $\textbf{E}\propto J_{s}\times \sigma_{s}$ \cite{Ando2}. Hence, the asymmetry is the resultant of the projection of spin polarization $\sigma_{s}$ in transverse and longitudinal geometries and the damping of SAW in the propagation direction (see Supplemental Material \cite{Suppl}). 

Magnitude of spin current generated via magnon-phonon coupling in our samples, can be estimated by \cite{Rojas}

\begin{equation}
\begin{split}\label{eq:LVolt}
J_{s}=\frac{V(\theta)}{\lambda_{IEE}wRsin\theta}
\end{split}
\end{equation}

where $\lambda_{IEE}$ is the IEE length, $w$ is the sample width and $R$ the electric sample resistance. For our Cu/Bi$_{2}$O$_{3}$ we have $\lambda_{IEE}=-0.17nm$\cite{Tsai}, $w=10\mu$m, $R=42.87\Omega$. We obtain a spin current J$_{s}=1.648\times10^{8}A/m^{2}$ for V$_{T}(60^{\circ})$ of Cu/Bi$_{2}$O$_{3}$ in Figure 4(a). This spin current density is comparable to commonly reported values for standard spin pumping driven by electromagnetic wave FMR \cite{Rojas, Zhang, Jamali}. Full angle dependence of spin current is available in the Supplemental Material \cite{Suppl}.

In summary, we demonstrated the feasibility of combining acoustic spin pumping via magnon - phonon coupling and spin to charge conversion via IEE at the Rashba like Cu(Ag)/Bi$_{2}$O$_{3}$ interfaces. Opposite signs of acoustic spin pumping signals elucidates the opposite spin configuration for Cu/Bi$_{2}$O$_{3}$ and Ag/Bi$_{2}$O$_{3}$ interfaces. Combination of full angle dependence of in plane magnetic field with transverse and longitudinal voltage detection allows to study the contributions of longitudinal and shear waves always present in Rayleigh type SAW. Such contributions are observed as asymmetries present in the magnetic field dependence of absorption and acoustic spin pumping measurements. The present study gives a systematic study of mechanical generation of spin current via magnon - phonon interaction. Alternatively, it has been proposed that mechanical generation of spin current can also be achieved by direct coupling to the atomic lattice, with no need for magnetic materials, external magnetic fields and spin orbit interaction \cite{Matsuo}. Experimental manifestation of this coupling has been recently observed by coupling SAW to Cu layer \cite{Nozaki}. Despite that our experimental scheme cannot rule out additional generation of spin current via direct SAW coupling to our nonmagnetic layers (Cu, Ag), we do not observe clear evidence of it in our power absorption measurements at $\theta= 90^{\circ}$ or $270^{\circ}$, where the magnetization in Ni is expected to be perpendicular to the vector of spin polarization induced by spin rotation coupling (Figure 2). This is the configuration for maximum signal reported in \cite{Nozaki}. Unambiguous determination of spin current generation by direct spin rotation coupling in Cu (Ag) may be achieved by optical characterization of spin accumulation without presence of magnetic materials \cite{Puebla}. 

\hfill \break
This work was supported by Grant-in-Aid for Scientific Research
on Innovative Area, “Nano Spin Conversion Science” (Grant
No. 26103002) and RIKEN Incentive Research Project Grant
No. FY2016. F.A. was supported by the Ministry of
Education, Culture, Sports, Science and Technology (MEXT)
Scholarship, Japan.



\begin{thebibliography}{1}

  \bibitem{Otani} Otani, Y., Shiraishi, M., Oiwa, A., Saitoh, E., and Murakami, S., Nature Physics (2017)

  \bibitem{Weiler1}  M. Weiler, L. Dreher, C. Heeg, H. Huebl, R. Gross, M. S. Brandt, and S. T. B. Goennenwein, PRL 106, 117601 (2011)

  \bibitem{Yaro} Yaroslav Tserkovnyak, Arne Brataas, and Gerrit E. W. Bauer, Phys. Rev. B 66, 224403 (2002)
  
\bibitem{Weiler2} M. Weiler, H. Huebl, F. S. Goerg, F. D. Czeschka, R. Gross, and S. T. B. Goennenwein, PRL 108, 176601 (2012)  
  
  \bibitem{Mosendz} O. Mosendz, V. Vlaminck, J. E. Pearson, F. Y. Fradin, G. E. W. Bauer, S. D. Bader, and A. Hoffmann, Phys. Rev. B 82, 214403 (2010)

\bibitem{Karube} S. Karube, K. Kondou, and Y. Otani, Appl. Phys. Express 9, 033001 (2016)

\bibitem {Rojas} J. C. Rojas-Sanchez, L. Vila, G. Desfonds, S. Gambarelli, J. P. Attane, J. M. De Teresa, C. Magen, and A. Fert, Nat. Commun. 4, 2944 (2013)

\bibitem{Ando} Y. Ando and M. Shiraishi, J. Phys. Soc. Jpn. 86, 011001 (2017)

\bibitem{Puebla} J. Puebla, F. Auvray, M. Xu, B. Rana, A. Albouy, H. Tsai, K. Kondou, G. Tatara, and Y. Otani, Appl. Phys. Lett. 111, 092402 (2017)

\bibitem{Kim} J. Kim, Y-T. Chen, S. Karube, S. Takahashi, K. Kondou, G. Tatara and Y. Otani, Phys. Rev. B 96, 140409(R) (2017).

\bibitem{Dreher} L. Dreher, M. Weiler, M. Pernpeintner, H. Huebl, R. Gross, M. S. Brandt, and S. T. B. Goennenwein, Phys. Rev. B 86, 134415 (2012)

\bibitem{Onose} R. Sasaki, Y. Nii, Y. Iguchi, and Y. Onose, Phys. Rev. B 95, 020407 (2017)

\bibitem{Nazar} Z. Nazarchuk, V. Skalskyi, O. Serhiyenko, Acoustic Emission: Methodology and Application, Fundation of Engineering Mechanics, Spinger  (2017)

\bibitem{Tsai} H. Tsai, S. Karube, K. Kondou, N. Yamaguchi, F. Ishii, Y. Otani, under consideration Phys. Rev. B (2018)

\bibitem{Azevedo} A. Azevedo, L. H. Vilela-Leao, R. L. Rodrıguez-Suarez, A. F. Lacerda Santos, and S. M. Rezende, Phys. Rev. B, vol. 83, 144402 (2011)

\bibitem{Ando2} K. Ando, S. Takahashi, J. Ieda, Y. Kajiwara, H. Nakayama, T. Yoshino, K. Harii, Y. Fujikawa, M. Matsuo, S. Maekawa, E. Saitoh, J. Appl. Phys. 109, 103913 (2013)

\bibitem{Suppl} See Supplemental Material at https://doi.org/10.1103/PhysRevB.97.180301 for full angular variation of spin current, details of fitting formulas for IEE voltage detection, XRD characterization and description of anisotropic distribution of charge potential. 

\bibitem{Zhang} W. Zhang, M. B. Jungfleisch, W. Jiang, J. E. Pearson, and A. Hoffmann, Journal of Applied Physics 117, 17C727 (2015)

\bibitem{Jamali} M. Jamali, J.S. Lee, J.S. Jeong, F. Mahfouz, Y. Lv, Z. Zhao, Branislav K. Nikolic, K. Andre Mkhoyan, N. Samarth, J-P. Wang, Nano Letters 15, 7126 (2015)

\bibitem{Matsuo} Mamoru Matsuo, Junichi Ieda, Kazuya Harii, Eiji Saitoh, and Sadamichi Maekawa, Phys. Rev. B 87 180402 (2013)

\bibitem{Nozaki} D. Kobayashi, T. Yoshikawa, M. Matsuo, R. Iguchi, S. Maekawa, E. Saitoh, and Y. Nozaki, Phys. Rev. Lett. 119, 077202 (2017) 

\end{thebibliography}
\end{document}